\documentclass[wcp]{jmlr_edit}

\usepackage{longtable}% for long tables

\usepackage{booktabs}
\usepackage{siunitx}
\usepackage{etoolbox}
\newtoggle{shorter}
\toggletrue{shorter}

\usepackage{algorithm}
\usepackage{algorithmic}
\usepackage{enumitem}
\usepackage{color}
\usepackage{multirow}
\definecolor{commentcolour}{RGB}{215,48,39}		% Comment in Red

\newcommand{\NAcell}{\multicolumn{2}{c}{\scriptsize$\mbox{N/A}$}}

\firstpageno{142}

\jmlrvolume{39}
\jmlryear{2014}
\jmlrworkshop{ACML 2014}

\title[Bibliographic Analysis with the Citation Network Topic Model]{Bibliographic Analysis with the\\ Citation Network Topic Model}

  \author{\Name{Kar Wai Lim} \Email{karwai.lim@anu.edu.au}\\
  \addr Australian National University, Canberra, Australia \\
  NICTA, Canberra, Australia
  \AND
  \Name{Wray Buntine} \Email{wray.buntine@monash.edu}\\
  \addr Monash University, Clayton, Australia
 }

\editors{Dinh Phung and Hang Li}

\begin{document}

\maketitle

\begin{abstract}
Bibliographic analysis considers author's research areas, the 
citation network and paper content among other things. In this 
paper, we combine these three in a topic model that produces a 
bibliographic model of authors, topics and documents using a 
non-parametric extension of a combination of the Poisson mixed-topic 
link model and the author-topic model. We propose a novel and 
efficient inference algorithm for the model to explore subsets of 
research publications from CiteSeer$^\mathrm{X}$. Our model 
demonstrates improved performance in both model fitting and 
a clustering task compared to several baselines.
\end{abstract}


\section{Introduction}

% \comment{Using AU spelling since this is for ACML}

Models of bibliographic data need to consider many kinds of 
information. Articles are usually accompanied by metadata, for 
example, authors, publication data, categories and time. Cited 
papers can also be available. When authors' topic preferences are 
modelled, we need to associate the document topic information 
somehow with the authors'. Jointly modelling text data with citation 
network information can be challenging for topic models, and the 
problem is confounded when also modelling author-topic relationships.

In this paper, we propose a topic model to jointly model authors' 
topic preferences, text content and the citation network. The model 
is a non-parametric extension of previous models discussed in 
Section~\ref{sec:relatedwork}. We derive a novel algorithm that 
allows the probability vectors in the model to be integrated out, 
using simple assumptions and approximations, which give Markov 
chain Monte Carlo (MCMC) inference via discrete sampling. 
Section~\ref{sec:model},~\ref{sec:likelihood} 
and~\ref{sec:inference} detail our model and its inference algorithm.
Applying our model on research publication data, we demonstrate the 
model's improved performance, on both model fitting and a clustering 
task, compared to baselines. We describe the datasets used in 
Section~\ref{sec:data} and report on experiments in 
Section~\ref{sec:experiment}. Additionally, we qualitatively analyse 
the inference results produced by our model. We find that the topics 
returned have high comprehensibility.

\section{Related Work}
\label{sec:relatedwork}
Latent Dirichlet Allocation (LDA)
%\citep{Blei:2003:LDA:944919.944937} 
is the simplest Bayesian topic model used in modelling text, which 
also allows easy learning of the model. \cite{TehJor2010a} proposed 
the {\em Hierarchical Dirichlet process} (HDP) LDA, which utilises 
the Dirichlet process (DP) as a non-parametric prior which allows a 
non-symmetric, arbitrary dimensional topic prior to be used. 
Furthermore, one can replace the Dirichlet prior on the word vectors
with the \textit{Pitman-Yor Process} 
\iftoggle{shorter}{(PYP)}{(PYP, also known as the two-parameter 
Poisson Dirichlet process)}
\citep{Teh:2006:HBL:1220175.1220299}, which models the power-law
of word frequency distributions in natural language
%\iftoggle{shorter}{}{\citep{goldwater2006interpolating} yielding
%significant improvement}
\citep{Sato:2010:TMP:1835804.1835890}.

Variants of LDA allow incorporating more aspects of a particular task
and here we consider authorship and citation information. 
The \textit{author-topic model} (ATM) 
\citep{Rosen-Zvi:2004:AMA:1036843.1036902} uses 
the authorship information to restrict topic options based on author.
Some recent work jointly models the document citation network 
and text content. 
This includes the  
\textit{relational topic model} \citep{chang2010hierarchical},
the \textit{Poisson mixed-topic link model} (PMTLM) 
\citep{ZhuYGM:2013} and
\textit{Link-PLSA-LDA} \citep{Nallapati:2008:JLT:1401890.1401957}.
An extensive review of these models can be found in 
\cite{ZhuYGM:2013}.
The \textit{Citation Author Topic} (CAT) 
model \citep{Tu:2010:CAT:1944566.1944711} models the 
author-author network on publications based on citations using an 
extension of the ATM.
Note that our work is different to CAT in 
that we model the author-document-citation network 
instead of author-author network.

The \textit{Topic-Link LDA} \citep{Liu:2009:TLJ:1553374.1553460} 
jointly models author and text by using the distance between the 
document and author topic vectors.
Similarly the Twitter-Network topic model \citep{Lim2013Twitter} 
models the author (``follower'') network based on author topic 
vectors, but using a Gaussian process to model the network.
Note that our work considers the author-document-citation of 
\cite{Liu:2009:TLJ:1553374.1553460} using the techniques developed 
in \cite{Lim2013Twitter}, but using the PMTLM of \cite{ZhuYGM:2013} 
to model the network which lets one integrate PYP
hierarchies with the PMTLM using efficient MCMC sampling.

There is also existing work on analysing the degree of authors' 
influence. On publication data, 
\cite{Kataria:2011:CST:2283696.2283777} and 
\cite{Mimno:2007:MDL:1255175.1255196} analyse influential authors 
with topic models.
While \cite{Weng:2010:TFT:1718487.1718520}, 
\cite{Tang:2009:SIA:1557019.1557108} and 
\cite{Liu:2010:MTI:1871437.1871467} use topic models to analyse 
users' influence on social media.

\section{Citation Network Topic Model}
\label{sec:model}

\begin{figure}[t]
	\begin{center}
	\centerline{\includegraphics[width=0.65\columnwidth]{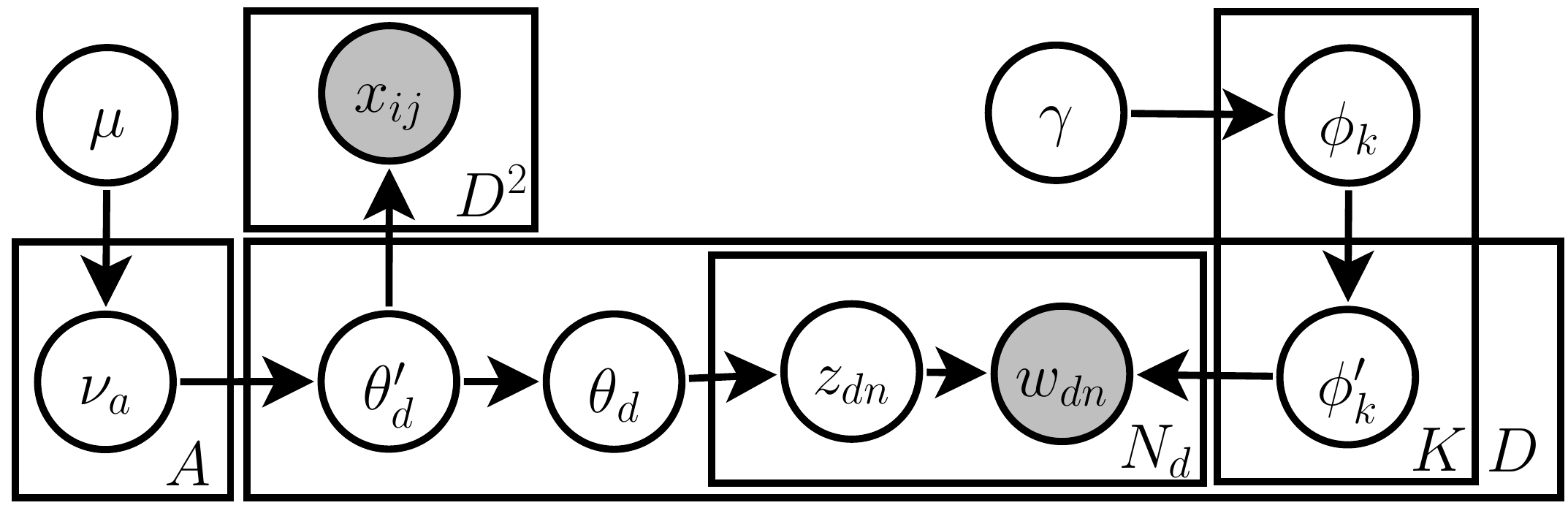} }
	\caption{Graphical model for CNTM. The box on the top left with 
		$D^2$ entries is the citation network on documents 
		represented as a Boolean matrix. The remainder is a 
		non-parametric author-topic model where the $A$ authors on 
		the left have topic vectors that influence the $D$ document 
		topic vectors. The $K$ topics, shown in the top right, have 
		bursty modelling following 
		\cite{Buntine:2014:ENT:2623330.2623691}.
	}
	\label{fig:poissonNetworkModel}
	\end{center}
	
\end{figure}

In this section, we propose a topic model that jointly model the 
\textit{text}, \textit{authors}, and the \textit{citation network} 
of research publications (documents). We name the topic model the 
Citation-Network Topic Model (CNTM). We first discuss the topic 
model part of CNTM where the citations are not considered, which
will be used for comparison later in Section~\ref{sec:experiment}.
The full graphical model for CNTM is displayed in 
Figure~\ref{fig:poissonNetworkModel}.
To clarify the notations used in this paper, \textit{variables that 
are without subscript represent a collection of variables of the 
same notation}. For example, $w_d$ would represent all the words in 
document $d$, that is, $w_d = \{w_{d1}, \dots, w_{dN_d}\}$ where 
$N_d$ is the number of words in document $d$; and $w$ represents all 
words in a corpus, $w=\{w_1, \dots, w_D\}$, where $D$ is the number 
of documents.

\subsection{Hierarchical Pitman-Yor Topic Model}
\label{subsec:topic_model}

The CNTM uses the \textit{Griffiths-Engen-McCloskey} (GEM) 
\citep{pitman1996some} distribution to generate probability vectors 
and the {\it Pitman-Yor process} 
(PYP)~\citep{Teh:2006:HBL:1220175.1220299} to generate probability 
vectors given another probability vector (called \textit{mean} or 
\textit{base} distribution). Both GEM and PYP are parameterised by a 
discount parameter $\alpha$ and a concentration parameter $\beta$.
PYP is additionally parameterised by a base distribution $H$, which 
is also the mean of the PYP. Note that the GEM distribution is 
equivalent to a PYP with a base distribution that generates an 
ordered integer label. 

In modelling authors, CNTM modifies the approach of the author-topic 
model \citep{Rosen-Zvi:2004:AMA:1036843.1036902}, which assumes that 
the words in a publication are equally attributed to the different 
authors. This is not reflected in practice since publications are 
often written more by the first author, excepting when the order is 
alphabetical. An approximation we make in this work is that the first
author is dominant. We could model the influence of each author on a 
publication, say, using a Dirichlet distribution, but we found that 
considering only the first author gives a simpler learning algorithm 
and cleaner results. 

IN CNTM, we first sample a root topic distribution $\mu$ with a GEM 
distribution, to act as a base distribution for the author-topic 
distributions $\nu_a$ for each author $a$:
\begin{align*}
\mu \sim \mathrm{GEM}(\alpha^\mu, \beta^\mu)~,
&&
\nu_a|\mu \sim \mathrm{PYP}(\alpha^{\nu_a}, \beta^{\nu_a}, \mu)~~~.
\end{align*}

\noindent
Given the first author $a_d$ of each publication $d$, we sample the 
document-topic prior $\theta'_d$ and the document-topic distribution 
$\theta_d$:
\begin{align*}
\theta'_d|a_d,\nu \sim 
\mathrm{PYP}(\alpha^{\theta'_d}, \beta^{\theta'_d}, \nu_{a_d})~,
&&
\theta_d|\theta'_d \sim 
\mathrm{PYP}(\alpha^{\theta_d}, \beta^{\theta_d}, \theta'_d)~~~.
\end{align*}
Note that instead of modelling a single document-topic distribution, 
we model a document-topic hierarchy with $\theta'$ and $\theta$.
The primed $\theta'$ represents the topics of the document in the 
context of the citation network. The unprimed $\theta$ represents 
the topics of the text, naturally related to $\theta'$ but not the 
same. Such modelling gives citation information a higher impact to 
counter the relatively low amount of citations compared to the text.
More details on the motivation of such modelling is presented in the
supplementary materials.

For the vocabulary side, we generate a background word distribution 
$\gamma$, where $H^\gamma$ is a discrete uniform vector of length 
$|\mathcal{V}|$ 
% that is, $H^\gamma = (\cdots, \frac{1}{|\mathcal{V}|}, \cdots)$. 
and $\mathcal{V}$ is the set of distinct word tokens observed.
Then, we sample a topic-word distribution $\phi_k$ for each topic 
$k$, with $\gamma$ as the base distribution:
\begin{align*}
\gamma \sim \mathrm{PYP}(\alpha^\gamma, \beta^\gamma, H^\gamma)~, 
&&
\phi_k|\gamma \sim 
\mathrm{PYP}(\alpha^{\phi_k}, \beta^{\phi_k}, \gamma)~~~.
\end{align*}
Modelling word burstiness \citep{Buntine:2014:ENT:2623330.2623691} 
is important since, as shown in Section~\ref{sec:data}, words in a 
document are likely to repeat in the document. This is addressed by 
making topics bursty, so each document only focuses on a subset of 
words in the topic. To generate $\phi'_{dk}$ for each topic $k$ in 
document $d$:
\begin{align*}
\phi'_{dk}|\phi_k \sim 
\mathrm{PYP}(\alpha^{\phi'_{dk}}, \beta^{\phi'_{dk}}, \phi_k)~~~.
\end{align*}

\noindent
Finally, for each word $w_{dn}$ in document $d$, we sample the 
corresponding topic assignment $z_{dn}$ from the document-topic 
distribution $\theta_d$, while the word $w_{dn}$ is sampled from the 
topic-word distribution $\phi'_d$ given $z_{dn}$.
\begin{align*}
z_{dn}|\theta_d \sim \mathrm{Discrete}(\theta_d)~,
&&
w_{dn}|z_{dn}, \phi'_d \sim \mathrm{Discrete}(\phi'_{dz_{dn}})~~~.
\end{align*}
Note that $w$ includes words from title and abstract, but not the 
full article of a publication. This is because title and abstract 
provide a good summary of a publication's topics, while the full 
article contains too much detail.

\subsection{Citation Network Poisson Model}
\label{subsec:network}

In CNTM, we assume that the citations are generated based on the 
topics relevant to the publications' using the degree-corrected 
version of the PMTLM \citep{ZhuYGM:2013}. Denoting $x_{ij}$ as the 
number of times document $i$ citing document $j$, we model $x_{ij}$
with a Poisson distribution with mean parameter $\lambda_{ij}$:
\begin{align}
x_{ij}|\lambda_{ij} \sim  \mathrm{Poisson}(\lambda_{ij})~,
&&
\textstyle
\lambda_{ij} = 
\lambda_i^+ \lambda_j^- \sum_k \lambda^T_k \theta'_{ik} 
\theta'_{jk}~~~.
\label{eq:lambda_ij}
\end{align}
Here, $\lambda_i^+$ is the propensity of document $i$ to cite and 
$\lambda_j^-$ represents the popularity of cited document $j$
and $\lambda^T_k$ scales the $k$-th topic. Hence, a citation from 
document $i$ to document $j$ is more likely when these documents are 
having relevant topics. The Poisson distribution\footnote{Note 
that Poisson distribution is similar to the Bernoulli distribution 
when the mean parameter is small.} is used instead of a Bernoulli
because it leads to dramatically reduced complexity in analysis.

\section{Model Representation and Posterior}
\label{sec:likelihood}

Before presenting the posterior used to develop the MCMC sampler, we 
briefly review  handling of hierarchical PYP models in 
Section~\ref{subsec:mhpyp}. We cannot provide an adequately 
detailed review in this paper, thus we present the main ideas.

\subsection{Modelling with Hierarchical PYPs}
\label{subsec:mhpyp}
The key to efficient Gibbs sampling with PYPs is to marginalise out 
the probability vectors (\textit{e.g.}\ topic distributions) in the 
model and record various associated counts instead, thus yielding a 
collapsed sampler. While a common approach here is to use the 
hierarchical Chinese Restaurant Process (CRP) of~\cite{TehJor2010a}, 
we use another representation that requires no dynamic memory and 
has better inference 
efficiency~\citep{Chen:2011:STC:2034063.2034095}. 

We denote $f(\mathcal{N})$ as the marginalised likelihood associated 
with the probability vector $\mathcal{N}$. Since the vector is 
marginalised out, the marginalised likelihood is in terms of --- 
using the CRP terminology --- the \textit{customer counts} 
$c^\mathcal{N} = (\cdots, c_k^\mathcal{N}, \cdots)$ and the 
\textit{table counts} $t^\mathcal{N} = (\cdots, t_k^\mathcal{N}, 
\cdots)$. The customer count $c_k^\mathcal{N}$ corresponds to the 
number of data points (\textit{e.g.}\ words) assigned to group $k$ 
(\textit{e.g.}\ topic) for variable $\mathcal{N}$. Here, the 
\textit{table counts} $t^\mathcal{N}$ represent the subset of 
$c^\mathcal{N}$ that gets passed up the hierarchy (as customers for 
the parent probability vector of $\mathcal{N}$). We also denote 
$C^\mathcal{N} = \sum_k c_k^\mathcal{N}$ as the total customer 
counts for node $\mathcal{N}$, and similarly, ${T}^\mathcal{N} = 
\sum_k t_k^\mathcal{N}$ is the total table counts. The marginalised 
likelihood is:
\begin{align}
f(\mathcal{N}) & = 
\frac{(\beta^\mathcal{N}|\alpha^\mathcal{N})_{{T}^\mathcal{N}}}
{(\beta^\mathcal{N})_{C^\mathcal{N}}} 
\prod_k 
S^{c^\mathcal{N}_k}_{{t}^\mathcal{N}_k, \alpha^\mathcal{N}} ~~~,
&&
\mathrm{for\ \ }
\mathcal{N} \sim 
\mathrm{PYP}(\alpha^\mathcal{N}, \beta^\mathcal{N}, \mathcal{P}) 
~~~.  
\label{eq:modularized_likelihood} 
\end{align} 
$S^x_{y,\alpha}$ is the generalised Stirling number; both $(x)_C$ 
and $(x|y)_C$ denote the Pochhammer symbol (rising factorial), see 
\cite{buntine2012bayesian} for details. Note the GEM distribution 
behaves like a PYP in which the table count $t_k^\mathcal{N}$ is 
always $1$ for non-zero $c_k^\mathcal{N}$.

The innovation of \cite{Chen:2011:STC:2034063.2034095} was to notice 
that sampling with Equation~\ref{eq:modularized_likelihood} directly 
led to poor performance due to inadequate mixing. They introduce a 
new Bernoulli \textit{indicator variable} $u^\mathcal{N}_k$ for each 
customer who has contributed a~``$+1$'' to ${c}^\mathcal{N}_k$. A 
value $u^\mathcal{N}_k=1$ indicates that the customer has opened a 
new table, which also means the customer has also contributed 
a~``$+1$'' to ${t}^\mathcal{N}_k$ and thus has been passed up the 
hierarchy to the parent variable $\mathcal{P}$.
The process repeats at the parent node because the ``$+1$'' to 
${t}^\mathcal{N}_k$ is inherited as a ``$+1$'' to 
${c}^\mathcal{P}_k$, and thus we now need to consider the value of 
$u^\mathcal{P}_k$. If $u^\mathcal{N}_k=0$ then a ``+1'' was not 
inherited and a corresponding $u^\mathcal{P}_k$ does not exist.
%\iftoggle{shorter}{}{Note 
%that the indicator variable $u^\mathcal{N}_k$ is not stored, that is, we simply `forget' who 
%contributed a table count and re-sample $u^\mathcal{N}_k$ as needed when decrementing:
%\begin{align}
%p(u^\mathcal{N}_k = 1) = t^{\mathcal{N}}_k / c^\mathcal{N}_k ~~~, &&
%p(u^\mathcal{N}_k = 0) = 1 - t^{\mathcal{N}}_k / c^\mathcal{N}_k ~~~.
%\end{align}
%}
The use of indicator variables has been empirically shown to lead to 
better mixing of the samplers.

Note that even though the probability vectors are integrated out and 
not explicitly stored, they can easily be estimated from the 
associated counts.  The probability vector 
$\mathcal{N}$ is estimated from the counts and parent 
probability vector $\mathcal{P}$ using standard CRP estimation:
\begin{align}
\mathcal{N} &= 
\left( \cdots , 
\frac{
(\alpha^\mathcal{N} T^\mathcal{N} + \beta^\mathcal{N}) \mathcal{P}_k 
+ c_k^\mathcal{N} - \alpha^\mathcal{N} T_k^\mathcal{N}
}{
\beta^\mathcal{N} + C^\mathcal{N}
} 
, \cdots \right) ~~~.
\label{eq:recover_vector}
\end{align}

\vspace{1mm}
\subsection{Likelihood for the Hierarchical PYP Topic Model}

We use bold face capital letters to denote the set of all relevant 
lower case variables, for example, 
$\mathbf{Z} = \{z_{11},\cdots,z_{DN_D}\}$ 
denotes the set of all topic assignments. Variables $\mathbf{W}, 
\mathbf{T}$ and $\mathbf{C}$ are similarly defined, that is, they 
denote the set of all words, table counts and customer counts 
respectively. Additionally, we denote $\mathbf{\zeta}$ as the set of 
all hyperparameters (such as the $\alpha$'s). With the probability 
vectors replaced by the counts, the likelihood of the topic model 
can be written --- in terms of $f(\cdot)$ --- as 
$p(\mathbf{Z}, \mathbf{W}, \mathbf{T}, \mathbf{C} | \mathbf{\zeta}) 
\propto$
\begin{align}
f(\mu) \Bigg(\prod_{a=1}^A f(\nu_a) \! \Bigg) 
\Bigg(\prod_{d=1}^D f(\theta'_d) \,
f(\theta_d)
\prod_{k=1}^K f(\phi'_{dk}) \! \Bigg)
\Bigg(\prod_{k=1}^K f(\phi_k) \! \Bigg)
f(\gamma) 
\Bigg( \! 
\prod_v \left(\frac{1}{|\mathcal{V}|}\right)^{t^\gamma_v} 
\Bigg) ~~~.
\label{eq:likelihood}
\end{align}
Note that the last term in Equation~\ref{eq:likelihood} corresponds 
to the parent probability vector of $\gamma$ (see 
Section~\ref{subsec:topic_model}), and $v$ indexes the unique word 
tokens in vocabulary set $\mathcal{V}$.

\subsection{Likelihood for the Citation Network Poisson Model}

For the citation network, the Poisson likelihood for each $x_{ij}$ 
\iftoggle{shorter}{uses the definition of $\lambda_{ij}$ in
Equation~\ref{eq:lambda_ij}.
Note that the term $x_{ij}!$ is dropped}{is given as
\begin{align}
p(x_{ij}|\lambda, \theta) = 
\frac{\lambda_{ij}^{x_{ij}}}{x_{ij}! \, e^{\lambda_{ij}}} 
= \left( 
\lambda_i^+ \lambda_j^- 
\sum_k \lambda^T_k \theta'_{ik} \theta'_{jk} 
\right)^{x_{ij}}
\exp \left(- \lambda_i^+ \lambda_j^-  
\sum_k \lambda^T_k \theta'_{ik} \theta'_{jk} \right)
~~~.
\label{eq:poisson_likelihood}
\end{align}
Note that the term $x_{ij}!$ is dropped in 
Equation~\ref{eq:poisson_likelihood}}
due to the limitation of the data that $x_{ij} \in \{0, 1\}$, thus 
$x_{ij}!$ is evaluated to $1$. With conditional independence of 
$x_{ij}$, the joint likelihood for the whole citation network
$\mathbf{X} = \{x_{11}, \cdots, x_{DD}\}$ can be written as
\begin{align*}
p(\mathbf{X}|\lambda, \theta') =
 \left( \prod_i 
 	(\lambda_i^{+})^{g^+_i} \, (\lambda_i^{-})^{g^-_i} 
 \right)
 \prod_{ij} 
 \left( 
 	\sum_k \lambda^T_k \theta'_{ik} \theta'_{jk} 
 \right)^{\! x_{ij}} 
 \exp\!\Bigg( 
 	- \sum_{ijk} \lambda_i^+ \lambda_j^- \lambda^T_k 
 	\theta'_{ik} \theta'_{jk} 
 \Bigg) ~~~,
\end{align*}
where $g^+_i$ is the number of citations for publication $i$, 
$g^+_i = \sum_j x_{ij}$, and $g^-_i$ is the number of times 
publication $i$ being cited, $g^-_i = \sum_j x_{ji}$. We also make a 
simplifying
assumption\footnote{Technically, defining $x_{ii}$ allows us to 
rewrite the joint likelihood into another form for efficient caching.
}
that $x_{ii} = 1$ for all documents $i$, that is, all publications 
are treated as self-cited.

In the next section, we demonstrate that our model representation 
gives rise to an intuitive sampling algorithm for learning the model.
We also show how the Poisson model integrates into the topic 
modelling framework.

\section{Inference Techniques}
\label{sec:inference}

Here, we derive the Markov chain Monte Carlo (MCMC) algorithms for 
learning the Citation Network Topic Model.
We first detail the Gibbs sampler for the topic model and then 
discuss the Metropolis-Hastings (MH) algorithm for the citation 
network. The full inference procedure is performed by alternating 
between the Gibbs sampler and the MH algorithm.

\subsection{Collapsed Gibbs Sampler for the Hierarchical PYP Topic Model}
\label{subsec:gibbs_sampler}

To jointly sample the words' topic and the associated counts in the 
CNTM, we use a collapsed Gibbs sampler designed for the PYP 
\citep{Chen:2011:STC:2034063.2034095}.
The concept of the sampler is analogous to LDA, which consists of 
decrementing the counts associated with a word, sampling the 
respective new topic assignment for the word, and incrementing the 
associated counts. 
Our collapsed Gibbs sampler is more complicated than LDA.
In particular, we have to consider the indicators $u^\mathcal{N}_k$ 
described in Section~\ref{subsec:mhpyp} operating on the hierarchy 
of PYPs.

The sampler proceeds by considering the latent variables associated 
with a given word $w_{dn}$.  
First, we decrement out the effects of the latent variables, the 
topic $z_{dn}=k$ and the chain of indicator variables 
$u^{\theta_d}_k, u^{\theta'_d}_k, u^{\nu_{a_d}}_k, u^{\mu}_k$
(where they exist).
After decrementing, we jointly sample a new topic $z_{dn}$
and the associated indicators (which contribute ``$+1$'' to counts) 
for word $w_{dn}$ from their joint conditional posterior 
distribution:
\begin{align}
p(z_{dn}, \mathbf{T}, \mathbf{C} | \mathbf{Z}^{-dn}, \mathbf{W}, 
\mathbf{T}^{-dn}, \mathbf{C}^{-dn}, \mathbf{\zeta})
= \frac{p(\mathbf{Z}, \mathbf{W}, \mathbf{T}, \mathbf{C} | 
\mathbf{\zeta})}{p(\mathbf{Z}^{-dn}, \mathbf{W}, \mathbf{T}^{-dn}, 
\mathbf{C}^{-dn} | \mathbf{\zeta})} 
~~~. 
\label{eq:conditional_posterior}
\end{align}
where the superscript $\Box^{-dn}$ indicates that the topic 
$z_{dn}$, indicators and the associated counts for word $w_{dn}$ are 
not observed in the respective sets, \textit{i.e.}\ the state after 
decrement. The modularised likelihood of 
Equation~\ref{eq:likelihood} allows the conditional posterior 
(Equation~\ref{eq:conditional_posterior}) to be computed easily, 
since it simplifies to ratios of likelihood $f(\cdot)$, which 
simplifies further as the counts differ by at most $1$ during 
sampling. For instance, the ratio of the Pochhammer symbols, 
$(x|y)_{C+1} / (x|y)_C$, simplifies to  $x+Cy$, while the ratio of 
Stirling numbers, such as $S^{y+1}_{x+1, \alpha}/S^{y}_{x, \alpha}$,
can be computed quickly via caching~\citep{buntine2012bayesian}.

Sampling a new $z_{dn} = k$ corresponds to incrementing the counts 
for variable $\theta_d$, that is, ``$+1$'' to $c^{\theta_d}_k$ and 
\textit{possibly} also ``$+1$'' to $t^{\theta_d}_k$.
If $t^{\theta_d}_k$ is incremented, then $c^{\theta'_d}_k$ will be 
incremented too but $t^{\theta'_d}_k$ may or may not be, as dictated 
by the sampled indicators $u_k$. The process is repeated until the 
root $\mu$, since $\mu$ is GEM distributed, incrementing $t^\mu_k$ 
is equivalent to sampling a \textit{new} topic, \textit{i.e.}\ the 
number of topics increase by $1$.
Procedure on the vocabulary side ($\phi$ \textit{etc}.) is similar.

\subsection{Metropolis-Hastings Algorithm for Citation Network}

\iftoggle{shorter}{We}{A naive MH algorithm can be proposed for 
learning the citation network. 
For instance, to sample the document topic distribution $\theta'_d$ 
given $\mathbf{X}$, we can propose a new $\theta'_d$ with a 
Dirichlet distribution given the parent vector $\nu_{a_d}$, and 
accept or reject the proposal following an MH scheme.
However, this algorithm requires the probability vectors 
($\theta'$, $\nu$, $\mu$) be stored explicitly and the collapsed 
Gibbs sampler in Section~\ref{subsec:gibbs_sampler} would be 
considerably more complicated.

Instead, we}
% end toggle
propose a novel MH algorithm that allows the probability vectors to 
remain integrated out, thus retaining the fast discrete sampling 
procedure for the PYP and GEM hierarchy, rather than, for instance, 
resorting to an expectation-maximisation (EM) algorithm or 
variational approach.
We introduce an \textit{auxiliary variable} $y_{ij}$, named 
\textit{citing topic}, to denote the topic that prompts publication 
$i$ to cite publication $j$.
To illustrate, for a \textit{biology} publication that cites a 
\textit{machine learning} publication for the learning technique, 
the citing topic would be `machine learning' instead of `biology'.
% When a citation is observed, \textit{i.e.}\ $x_{ij} = 1$, we model the
From Equation~\ref{eq:lambda_ij},
a citing topic $y_{ij}$ is jointly Poisson with $x_{ij}$:
\begin{align}
x_{ij}, y_{ij} = k | \lambda, \theta'
\sim \mathrm{Poisson}
 \left(\lambda^+_i \lambda^-_j \lambda^T_k \theta'_{ik} \theta'_{jk} \right)~.
\label{eq:citing_topic}
\end{align}
Incorporating $\mathbf{Y}$, the set of all $y_{ij}$, 
we rewrite the citation network likelihood as
\begin{align*}
p(\mathbf{X},\mathbf{Y}|\lambda, \theta') \propto 
\prod_i (\lambda_i^+)^{g^+_i} (\lambda_i^-)^{g^-_i} 
\prod_k \left(\lambda_k^T\right)^{\frac{1}{2}\sum_i h_{ik}} \!
\prod_{ik} {\theta'_{ik}}^{h_{ik}}
\, \exp \! \Bigg(- 
\sum_{ij} \lambda_i^+ \lambda_j^-  \lambda^T_{y_{ij}} 
\theta'_{i{y_{ij}}} \theta'_{j{y_{ij}}} \Bigg)
\end{align*}
where $h_{ik}=\sum_j x_{ij}I(y_{ij}=k)+\sum_j x_{ji}I(y_{ji}=k)$ 
is the number of connections publication $i$ made due to topic $k$.

To integrate out $\theta'$, we note the term 
${\theta'_{ik}}^{h_{ik}}$ appears like a multinomial likelihood, so 
we absorb them into the likelihood for 
$p(\mathbf{Z}, \mathbf{W}, \mathbf{T}, \mathbf{C} | \mathbf{\zeta})$ 
where they correspond to additional counts for $c^{\theta'_i}$, with 
$h_{ik}$ added to $c^{\theta'_i}_k$. To disambiguate the source of 
the counts, we will refer these customer counts contributed by 
$x_{ij}$ as \textit{network counts}, and denote the augmented counts 
($\mathbf{C}$ plus network counts) as $\mathbf{C}^+$.
For the exponential term, we use Delta method approximation,
$\int f(\theta)\,\exp(-g(\theta))\,\mathrm{d}\theta
\approx \exp(-g(\hat\theta)) \int f(\theta)\,\mathrm{d}\theta$,
where $\hat\theta$ is the expected value according to a distribution
proportional to $f(\theta)$. This approximation is reasonable as 
long as the terms in the exponential are small (see supplementary 
material). The approximate full posterior of CNTM can then be 
written as 
$p(\mathbf{Z}, \mathbf{W}, \mathbf{T}, \mathbf{C}, \mathbf{X}, 
\mathbf{Y}|\lambda,\mathbf{\zeta}) \approx$
\begin{align}
p(\mathbf{Z}, \mathbf{W}, \mathbf{T}, \mathbf{C}^+ | \mathbf{\zeta})
\prod_i (\lambda_i^+)^{g^+_i} (\lambda_i^-)^{g^-_i} 
\prod_k \left(\lambda_k^T\right)^{\frac{1}{2}\sum_i h_{ik}}
\, \exp \! 
\Bigg(- \sum_{ij} \lambda_i^+ \lambda_j^-  \lambda^T_{y_{ij}} 
\hat\theta'_{iy_{ij}} \hat\theta'_{jy_{ij}} 
\Bigg)
\label{eq:approx_network_likelihood}
\end{align}

The MH algorithm can be summarised in three steps: 
estimate the document topic prior $\theta'$, propose a new citing 
topic $y_{ij}$ from Equation~\ref{eq:citing_topic}, and accept or 
reject the proposed $y_{ij}$ following an MH scheme with 
Equation~\ref{eq:approx_network_likelihood}.
We present the details of the MH sampler in the supplementary 
material. Note that the MH algorithm is similar to the collapsed 
Gibbs sampler, where we decrement the counts, sample a new state and 
update the counts. Since all probability vectors are represented as 
counts, we do not need to deal with their vector form in the 
collapsed Gibbs sampler. Additionally, our MH algorithm is intuitive 
and simple to implement. Like the words in a document, each citation 
is assigned a topic, hence the words and citations can be thought as 
voting to determine a documents' topic.

\subsection{Hyperparameter Sampling}
\label{subsec:hyperparameter_sampling}

Hyperparameter sampling for the priors are 
important~\citep{WallachPrior2009}.
In our inference algorithm, we sample the concentration parameters 
$\beta$ of all PYPs with an auxiliary variable sampler 
\citep{Teh06abayesian},\footnote{We 
outline the hyperparameter sampling for concentration parameters in 
the supplementary material.} 
but leaving the discount parameters $\alpha$ fixed. We do not sample 
the $\alpha$ due to the coupling of the parameter with the Stirling 
numbers cache.

\iftoggle{shorter}{}{Assuming each $\beta^\mathcal{N}$ has a Gamma distributed hyperprior with shape $\tau_0$ and rate 
$\tau_1$, we first sample the auxiliary variables $\xi$ and $\psi_j$ for 
$j \in \{0, T^\mathcal{N} -1 \}$:
\begin{align*}
\xi | \beta^\mathcal{N} \sim \mathrm{Beta}(C^\mathcal{N}, \beta^\mathcal{N}) ~,
&&
\psi_j | \alpha^\mathcal{N}, \beta^\mathcal{N} \sim \mathrm{Bernoulli}\left(\frac{\beta^\mathcal{N}}{\beta^\mathcal{N}+j\alpha^\mathcal{N}}\right) ~~~.
\end{align*}
We then sample a new $\beta'{^\mathcal{N}}$ from the conditional posterior given the auxiliary 
variables:
\begin{align*}
\beta'{^\mathcal{N}} | \xi, \psi \sim \mathrm{Gamma}\left(\tau_0 + \textstyle{ \sum_j } \psi_j, \tau_1 - \log(1 - \xi)\right) ~~~.
\end{align*}}
% end toggle

In addition to the PYP hyperparameters, we also sample $\lambda^+$, 
$\lambda^-$ and $\lambda^T$ with a Gibbs sampler.
We let the hyperpriors for $\lambda^+$, $\lambda^-$ and $\lambda^T$ 
to be Gamma distributed with shape $\epsilon_0$ and rate 
$\epsilon_1$. With the conjugate Gamma prior, the posteriors for 
$\lambda^+_i$, $\lambda^-_i$ and $\lambda^T_k$ are also Gamma 
distributed, so they can be sampled directly.
\begin{align*}
(\lambda^+_i|\mathbf{X}, \lambda^-, \lambda^T \theta') 
& \sim 
\textstyle
\mathrm{Gamma}\left( \epsilon_0 + g^+_i, 
\epsilon_1 + \sum_k \lambda^T_k \theta'_{ik} 
\sum_j \lambda^-_j \theta'_{jk} \right) ~~~,
\\
(\lambda^-_i|\mathbf{X}, \lambda^+, \lambda^T \theta') 
& \sim 
\textstyle
\mathrm{Gamma}\left( \epsilon_0 + g^-_i, 
\epsilon_1 + \sum_k \lambda^T_k \theta'_{ik} 
\sum_j \lambda^+_j \theta'_{jk} \right) ~~~,
\\
(\lambda^T_k|\mathbf{X}, \mathbf{Y}, \lambda^+, \lambda^-, \theta') 
& \sim 
\textstyle
\mathrm{Gamma}\left( \epsilon_0 + \frac{1}{2} \sum_i h_{ik}, 
\epsilon_1 + \lambda^T_k (\sum_j \lambda^+_j \theta'_{jk}) 
(\sum_j \lambda^-_j \theta'_{jk}) \right) ~~~.
\end{align*}
In this paper, we apply vague priors to the hyperpriors by setting 
\iftoggle{shorter}{$\epsilon_0 = \epsilon_1 = 1$.}{$\tau_0 = \tau_1 = \epsilon_0 = \epsilon_1 = 1$.}

We summarise the full inference algorithm for the CNTM in 
Algorithm~\ref{alg:gibbs}.

\begin{algorithm}[ht!]
	\caption{Inference Algorithm for the Citation Network Topic Model}
	\label{alg:gibbs}
	\begin{enumerate}[itemindent=-10pt, itemsep=1pt]
		\item \parbox[t]{\dimexpr\linewidth+10pt}{
			Initialise the model by assigning a random topic 
			assignment $z_{dn}$ to each word $w_{dn}$ and 
			constructing the relevant customer counts 
			$c^\mathcal{N}_k$ and table counts $t^\mathcal{N}_k$ for 
			all variables $\mathcal{N}$.
		}
		\item For each word $w_{dn}$ in each document $d$:
		\begin{enumerate}[itemindent=-10pt, noitemsep, nolistsep,
			label=\roman{*}., ref=(\roman{*})]
  			\item Decrement the counts associated with $z_{dn}$ and 
  			$w_{dn}$~.
		    \item Blocked sample a new topic $z_{dn}$ and associated 
		    	$\mathbf{T}$ and $\mathbf{C}$ from 
		    	Equation~\ref{eq:conditional_posterior}.
		\end{enumerate}
		\item For each citation $x_{ij}$:
		\begin{enumerate}[itemindent=-10pt, noitemsep, nolistsep, 
			label=\roman{*}., ref=(\roman{*})]
  			\item Decrement the network counts associated with 
  			$x_{ij}$ and $y_{ij}$~.
		    \item Sample a new citing topic $y_{ij}$ from the joint 
		    posterior of Equation~\ref{eq:citing_topic}.
		    \item Accept or reject the sampled $y_{ij}$ with an MH 
		    scheme using Equation~\ref{eq:approx_network_likelihood}.
		\end{enumerate}
		\item Update the hyperparameters $\beta$, $\lambda^+$,
			$\lambda^-$ and $\lambda^T$.
		\item \parbox[t]{\dimexpr\linewidth+10pt}{Repeat steps 2-4 
			until the model converges or a fix number of iterations 
			reached.
		}
	\end{enumerate}
	
\end{algorithm}

\section{Data}
\label{sec:data}

We perform our experiments on subsets of CiteSeer$^\mathrm{X}$ 
data\footnote{\url{http://citeseerx.ist.psu.edu/}} which consists of 
scientific publications. Each publication from CiteSeer$^\mathrm{X}$ 
is accompanied by {\it title}, {\it abstract}, {\it keywords}, 
{\it authors}, {\it citations} and other metadata. We prepare three 
publication datasets from CiteSeer$^\mathrm{X}$ for evaluations.
The first dataset corresponds to Machine Learning (ML) publications, 
which are queried from CiteSeer$^\mathrm{X}$ using the keywords from 
Microsoft Academic 
Search.\footnote{\url{http://academic.research.microsoft.com/}}
The ML dataset contains 139,227 publications.

Our second dataset corresponds to publications from 10 distinct 
research areas: {\it agriculture}, {\it archaeology}, {\it biology}, 
{\it computer science}, {\it financial economics}, {\it industrial 
engineering}, {\it material science}, {\it petroleum chemistry}, 
{\it physics} and {\it social science}.
The query words for these 10 disciplines are chosen such that the 
publications form distinct clusters. We name this dataset M10 
(Multidisciplinary 10 classes), which is made of 10,310 publications.
For the third dataset, we query publications from both arts and 
science disciplines. Arts publications are made of \textit{history} 
and \textit{religion} publications, while the science publications 
contain \textit{physics}, \textit{chemistry} and \textit{biology} 
researches. This dataset consists of 18,720 publications and is 
named AvS (Arts versus Science) in this paper.

The keywords used to create the datasets are obtained from Microsoft 
Academic Search, and are listed in the supplementary material.
For the clustering evaluation in Section~\ref{subsubsec:clustering}, 
we treat the query categories as the ground truth. However, 
publications that span multiple disciplines can be problematic for 
clustering evaluation, hence we simply remove the publications that 
satisfy the queries from more than one discipline. Nonetheless, the 
labels are inherently noisy. The metadata for the publications can 
also be noisy, for instance, the {\it authors} field may sometimes 
display publication's keywords instead of the authors, publication 
title is sometimes an URL, and table of contents can be mistakenly 
parsed as the abstract. We discuss our treatments to these issues in 
Section~\ref{subsec:preprocessing}. We also note that non-English 
publications are discarded using 
{\tt langid.py}~\citep{Lui:2012:LOL:2390470.2390475}.

In addition to the manually queried datasets, we also make use of 
existing datasets from LINQS 
\citep{sen:aimag08}\footnote{\url{http://linqs.cs.umd.edu/projects/projects/lbc/}} 
to facilitate comparison with existing work. In particular, we use 
their CiteSeer, Cora and PubMed datasets. Their CiteSeer data 
consists of Artificial Intelligence publications and hence we name 
the dataset AI in this paper. Although these datasets are small, 
they are fully labelled and thus useful for clustering evaluation.
However, they do not come with additional metadata such as the 
authors. Note that the AI and Cora datasets are presented as 
Boolean matrices, \textit{i.e.}\ the word counts information is lost 
and all words in a document are assumed to occur only once.
Although this representation is less useful for topic modelling, we 
still use them for the sake of comparison. Also note that the word 
counts were converted to TF-IDF in the PubMed dataset, so we recover 
the word counts using a reasonable assumption, see supplementary 
material for the recovery process. In Table~\ref{tbl:datasets}, we 
present a summary of the datasets used in this paper. 

\begin{table}[t]
	\small
	\centering
	\begin{tabular}
	{
	l
	S[table-format=6.0]
	S[table-format=7.0]
	S[table-format=5.0]
	S[table-format=4.0]
	S[table-format=2.1]
	S[table-format=2.1]
	}
    \toprule
	Datasets & {Publications} & {Citations} & {Authors} & 
	{Vocabulary} & {Words/Doc} & \% {Repeat} \\
	\midrule
	1. ML     & 139227 & 1105462 & 43643 & 8322 & 59.4 &  23.3 \\
	2. M10    &  10310 &   77222 &  6423 & 2956 & 57.8 &  24.3 \\
	3. AvS    &  18720 &   54601 & 11898 & 4770 & 58.9 &  17.0 \\
	4. AI     &   3312 &    4608 & {$-$} & 3703 & 31.8 & {$-$} \\
	5. Cora   &   2708 &    5429 & {$-$} & 1433 & 18.2 & {$-$} \\
	6. PubMed &  19717 &   44335 & {$-$} & 4209 & 67.6 &  40.1 \\
	\bottomrule
	\end{tabular}
	\caption{Summary of the datasets used in the paper, showing the 
		number of publications, citations, 	authors, unique word 
		tokens, the average number of words in each document, and 
		the last column is the average percentage of unique words 
		repeated in a document. Note: author information is not 
		available on the last three datasets. 
	}
	\label{tbl:datasets}
	
\end{table}

\subsection{Data Noise Removal}
\label{subsec:preprocessing}

Here, we briefly discuss the steps taken in cleansing the noise from 
the CiteSeer$^\mathrm{X}$ datasets (ML, M10 and AvS).
Note that the {\it keywords} field in the publications are often 
empty and are sometimes noisy, that is, they contain irrelevant 
information such as section heading and title, which makes the 
keywords unreliable source of information as categories.
Instead, we simply treat the keywords as part of the abstracts.
We also remove the URLs from the data since they do not provide 
any additional useful information.

Moreover, the author information is not consistently presented in 
CiteSeer$^\mathrm{X}$. Some of the authors are shown with full name, 
some with first name initialised, while some others are prefixed 
with title (Prof, Dr.\ {\it etc.}). We thus standardise the author 
information by removing all title from the authors, initialising all 
first names and discarding the middle names. Although 
standardisation allows us to match up the authors, it does not solve 
the problem that different authors who have the same initial and 
last name are treated as a single author. For example, both Bruce 
Lee and Brett Lee are standardised to B Lee.  Note this 
corresponds to a whole research problem 
\citep{Han:2004:TSL:996350.996419, Han:2005:NDA:1065385.1065462} and 
hence not addressed in this paper. Occasionally, institutions are 
mistakenly treated as authors in CiteSeer$^\mathrm{X}$ data, 
example includes {\it American Mathematical Society} and 
{\it Technische Universit\"{a}t M\"{u}nchen}. In this case, we 
simply remove the incorrect authors using a list of exclusion 
words\footnote{The list of exclusion words is presented in the 
supplementary material.} 
for authors.

\subsection{Text Preprocessing}

Here, we discuss the preprocessing pipeline adopted for the 
\textit{queried} datasets (LINQS data were already processed).
First, since publication text contains many technical terms that are 
made of multiple words, we tokenise the text using phrases (or 
collocations) instead of \textit{unigram} words. Thus, phrases like 
{\it decision tree} are treated as single token rather than two 
distinct words. The phrases are extracted from the respective 
datasets using 
{\tt LingPipe}.\footnote{\url{http://alias-i.com/lingpipe/}}
In this paper, we use the word {\it words} to mean both unigram 
words and phrases.

We then change all the words to lower case and filter out certain 
words. Words that are removed are {\it stop words}, common words and 
rare words. More specifically, we use the stop words list from
{\tt MALLET},\footnote{\url{http://mallet.cs.umass.edu/}}
we define common words as words that appear in more than 18\% of the 
publications, and rare words are words that occur less than 50 times 
in each dataset.  Note that the threshold are determined by 
inspecting the words removed. Finally, the tokenised words are 
stored as arrays of integers. We also split the datasets to 90\% 
training set for training the topic models, and 10\% test set for 
evaluations detailed in Section~\ref{sec:experiment}.

\section{Experiments}
\label{sec:experiment}

In this section, we describe experiments that compare the CNTM 
against several baseline topic models. The baselines are 
HDP-LDA with burstiness~\citep{Buntine:2014:ENT:2623330.2623691}, 
a non-parametric extension of the ATM, the Poisson mixed-topic link 
model (PMTLM) \citep{ZhuYGM:2013} and the CNTM without the citation 
network. We evaluate these models quantitatively with 
goodness-of-fit and clustering measures. We qualitatively analyse 
the topics produced and perform topic analysis on the authors. 
Additionally, we experiment on merging authors who have low number 
of publications and grouping them based on categories. This gives us 
a semi-supervised topic modelling in which some labels are known for 
authors who do not publish much. Finally, we present a discussion on 
the algorithm running time and convergence analysis in the 
supplementary material.

In the following experiments, we initialise the concentration 
parameters $\beta$ of all PYPs to $0.1$, noting that the 
hyperparameters are updated automatically. We set the discount 
parameters $\alpha$ to $0.7$ for all PYPs corresponding to the 
``\textit{word}'' side of the CNTM (\textit{i.e.}\ $\gamma$, $\phi$, 
$\phi'$). This is to induce power-law behaviour on the word 
distributions. We simply fix the $\alpha$ to $0.01$ for all other 
PYPs.  Note that the number of topics grow with data in 
non-parametric topic modelling. To prevent the learned topics to be 
too fine-grained, we set a limit to the maximum number of 
topics that can be learned. In particular, we set the number of 
topics cap to 20 for the ML dataset, 50 for M10 and 30 for the AvS 
dataset. For all the topic models, our experiments find that the 
number of topics always converges to the cap. For AI, Cora and 
PubMed datasets, we \textit{fix} the number of topics to 6, 7 and 3 
respectively simply for comparison against PMTLM.

When training the topic models, we run the inference algorithm for 
2,000 iterations. For the CNTM, the MH algorithm for the citation 
network is performed after 1,000 iterations, this is so the
topics can be learned first. This gives a faster learning algorithm 
and also allows us to assess the ``\textit{value-added}'' by 
the citation network to 
topic modelling.\footnote{This is elaborated further in the 
supplementary material with likelihood comparison.}
We repeat each experiment five times to reduce the estimation error 
of the evaluation measures.

\subsection{Quantitative Results}
\subsubsection{Goodness-of-fit and Perplexity}

Perplexity is a popular metric used to evaluate the goodness-of-fit 
of a topic model. Perplexity is negatively related to the likelihood 
of the observed words given the model, and lower is better.
Perplexity, estimated using document completion, is given as:
\begin{align*}
\mathrm{perplexity}(\mathbf{W}) = 
\exp\left(
-\frac{
\sum_{d=1}^D \sum_{n=1}^{N_d} \log p(w_{dn}|\theta_d, \phi)}
{\sum_{d=1}^D N_d}
\right) ~~~,
\end{align*}
where $p(w_{dn}|\theta_d, \phi)$ is obtained by summing over all 
possible topics:
\begin{align*}
%\textstyle
p(w_{dn}|\theta_d, \phi) = 
\sum_k p(w_{dn}|z_{dn}=k, \phi_k) \, p(z_{dn}=k|\theta_d) 
= \sum_k \phi_{kw_{dn}} \theta_{dk} ~~~.
\end{align*}

\noindent
The topic distribution $\theta$ is unknown for the test documents.
Instead of using half of the text in the test set to estimate 
$\theta$, which is a standard practice, we used only the words from 
the title to estimate $\theta$. One of the reasons behind this is 
that although title is usually short, it is a good indicator of 
topic. Moreover, using only the title allows more words to be used 
to calculate the perplexity. The technical details on estimating 
$\theta$ is presented in the supplementary material. Note that the 
perplexity estimate is unbiased since the data used in estimating 
$\theta$ is not used for evaluation.

We present the perplexity result in Table~\ref{tbl:perplexity}, 
which clearly shows the 
significantly\footnote{In this paper, significance is quantified at 
5\% significance level.} 
better performance of CNTM against the baselines. Inclusion of 
citation information also provides significant improvement for model 
fitting, as shown in the comparison of CNTM with and without network 
component.

\begin{table}[t!]
	\small
	\centering
	\begin{tabular}{rr@{\,\tiny$\pm$\,}lr@{\,\tiny$\pm$\,}lr@{\,\tiny$\pm$\,}lr@{\,\tiny$\pm$\,}l}
	\toprule
	& 
	\multicolumn{4}{c}{ML} & 
	\multicolumn{4}{c}{M10} \\
%    \hline
	& 
	\multicolumn{2}{c}{Train} & 
	\multicolumn{2}{c}{Test} & 
	\multicolumn{2}{c}{Train} & 
	\multicolumn{2}{c}{Test} \\
	\midrule
	Bursty HDP-LDA & 
	$ 4904.24 $ & {\tiny $ 71.34 $} & $ 4992.94 $ & {\tiny $ 65.57 $} & 
	$ 1959.36 $ & {\tiny $ 32.77 $} & $ 2265.18 $ & {\tiny $ 68.19 $} \\ 
%	\hline
	Non-parametric ATM & 
	$ 2238.19 $ & {\tiny $ 12.22 $} & $ 2460.28 $ & {\tiny $ 11.34 $} & 
	$ 1562.85 $ & {\tiny $ 18.11 $} & $ 1814.03 $ & {\tiny $ 23.18 $}  \\ 
%	\hline
	CNTM w/o network & 
	$ 1918.21 $ & {\tiny $ 4.31 $} & $ 2057.61 $ & {\tiny $ 3.56 $} & 
	$ 912.69 $ & {\tiny $ 10.94 $} & $ 1186.11 $ & {\tiny $ 8.32 $} \\
%	\hline
	CNTM w network & 
	$ \textbf{1851.82} $ & {\tiny $ 8.50 $} & $ \textbf{1990.78} $ & {\tiny $ 11.36 $} & 
	$ \textbf{824.04} $ & {\tiny $ 11.96 $} & $ \textbf{1048.33} $ & {\tiny $ 21.39 $} \\
	\bottomrule
	\end{tabular}
	\caption{Perplexity for the train and test documents on ML and M10, lower is better.}
	\label{tbl:perplexity}
	
\end{table}

\subsubsection{Document Clustering}
\label{subsubsec:clustering}

Next, we evaluate the clustering ability of the topic models.
Recall that topic models assign a topic to each word in a document, 
essentially performing a soft clustering in which the membership is 
given by the document-topic distribution $\theta$.
For the following evaluation, we convert the soft clustering to hard 
clustering by choosing a topic that best represents the documents, 
hereafter called the \textit{dominant topic}. The dominant topic 
corresponds to the topic that has the highest probability in a topic 
distribution $\mathcal{N}$.

As mentioned in Section~\ref{sec:data}, we assume the ground truth 
classes correspond to the query categories used in creating the 
datasets. We evaluate the clustering performance with purity and 
normalised mutual information 
(NMI)\footnote{Note that the NMI in \cite{ZhuYGM:2013} is slightly 
different to ours, we use the definition in 
\cite{Manning:2008:IIR:1394399}. This \textit{penalises} our 
NMI result when compared against the result in \cite{ZhuYGM:2013}
since our normalising term will always be equal or greater than that 
of \cite{ZhuYGM:2013}.
}
\citep{Manning:2008:IIR:1394399}. 
Purity is a simple clustering measure which can be interpreted as 
the proportion of documents correctly clustered.
For ground truth classes $\mathcal{S} = \{s_1, \dots, s_J\}$ and 
obtained clusters $\mathcal{R} = \{r_1, \dots, r_K\}$, the purity 
and NMI are computed as 
\begin{align*}
\mathrm{purity}(\mathcal{S}, \mathcal{R}) = 
\frac{1}{D} \sum_k \max_j | r_k \cap s_j | ~~~,
&& 
\mathrm{NMI}(\mathcal{S}, \mathcal{R}) = 
\frac{2 \, I(\mathcal{S}; 
\mathcal{R})}{H(\mathcal{S})+H(\mathcal{R})} ~~~,
\end{align*}
where $I(\mathcal{S}; \mathcal{R})$ denotes the mutual information 
and $H(\cdot)$ denotes the entropy:
\begin{align*}
I(\mathcal{S}; \mathcal{R}) = 
\sum_{k,\,j} 
\frac{| r_k \cap s_j |}{D} 
\log_2 \frac{D |r_k \cap s_j|}{| r_k | | s_j |} ~~~,
&& 
H(\mathcal{R}) = - \sum_k \frac{|r_k|}{D} \log_2 \frac{|r_k|}{D} 
~~~.
\end{align*}

The clustering results are presented in Table~\ref{tbl:clustering} 
and Table~\ref{tbl:clustering_LINQS}. We can see that the CNTM 
greatly outperforms the PMTLM in NMI evaluation.  Note that for a 
fair comparison against PMTLM, the experiments on the AI, Cora and 
PubMed datasets are evaluated with a 10-fold cross validation.
Additionally, we would like to point out that since no author 
information is provided on these 3 datasets, the CNTM becomes a 
variant of HDP-LDA, but with PYP instead of DP. We find that the 
clustering performance of CNTM with or without network is similar in 
Table~\ref{tbl:clustering_LINQS}. This is likely because the 
publications in each datasets are highly related to one 
another,\footnote{See the list of category labels of these datasets 
in supplementary material.} and thus the citation information is 
not discriminating enough for clustering.

\begin{table}[tb!]
	\small
	\centering
	\begin{tabular}{rr@{\,\tiny$\pm$\,}lr@{\,\tiny$\pm$\,}lr@{\,\tiny$\pm$\,}lr@{\,\tiny$\pm$\,}l}
	\toprule
	& \multicolumn{4}{c}{M10} 
	& \multicolumn{4}{c}{AvS} 
	\\
%    \hline
	& \multicolumn{2}{c}{Purity} 
	& \multicolumn{2}{c}{NMI} 
	& \multicolumn{2}{c}{Purity} 
	& \multicolumn{2}{c}{NMI} 
	\\
	\midrule
	Bursty HDP-LDA 
	& $ 0.66 $ & {\tiny $ 0.02 $} 
	& $ 0.67 $ & {\tiny $ 0.01 $} 
	& $ \textbf{0.75} $ & {\tiny $ 0.03 $} 
	& $ \textbf{0.66} $ & {\tiny $ 0.01 $}
	\\ 
%	\hline
	Non-parametric ATM 
	& $ 0.58 $ & {\tiny $ 0.01 $} 
	& $ 0.63 $ & {\tiny $ 0.00 $} 
	& $ 0.69 $ & {\tiny $ 0.02 $} 
	& $ 0.64 $ & {\tiny $ 0.01 $}
	\\ 
%	\hline
	CNTM w/o network
	& $ 0.61 $ & {\tiny $ 0.04 $}  
	& $ 0.67 $ & {\tiny $ 0.01 $} 
	& $ 0.72 $ & {\tiny $ 0.03 $} 
	& $ \textbf{0.66} $ & {\tiny $ 0.01 $}
	\\ 
	CNTM w network
	& $ \textbf{0.67} $ & {\tiny $ 0.03 $} 
	& $ \textbf{0.69} $ & {\tiny $ 0.02 $} 
	& $ 0.72 $ & {\tiny $ 0.01 $} 
	& $ \textbf{0.66} $ & {\tiny $ 0.00 $}
	\\ 
	\bottomrule
	\end{tabular}
	\caption{Comparison of clustering performance on the M10 and AvS dataset.}
	\label{tbl:clustering}
	\vspace{2mm}
\end{table}

\begin{table}[htb!]
	\small
	\centering
	\begin{tabular}{rr@{\,\tiny$\pm$\,}lr@{\,\tiny$\pm$\,}lr@{\,\tiny$\pm$\,}lr@{\,\tiny$\pm$\,}lr@{\,\tiny$\pm$\,}lr@{\,\tiny$\pm$\,}l}
	\toprule
	& \multicolumn{4}{c}{AI} 
	& \multicolumn{4}{c}{Cora} 
	& \multicolumn{4}{c}{PubMed} \\
%    \hline
	& \multicolumn{2}{c}{Purity} & \multicolumn{2}{c}{NMI} 
	& \multicolumn{2}{c}{Purity} & \multicolumn{2}{c}{NMI} 
	& \multicolumn{2}{c}{Purity} & \multicolumn{2}{c}{NMI} \\
	\midrule
	PMTLM*
	& \NAcell 
	& \multicolumn{1}{c}{$ 0.51$\:} &
	& \NAcell 
	& \multicolumn{1}{c}{$ 0.41 $\:} &
	& \NAcell 
	& \multicolumn{1}{c}{$ 0.27 $\;} & \\ 
	CNTM w/o network
	& $ \textbf{0.51} $ & {\tiny $ 0.07 $}
	& $ \textbf{0.67} $ & {\tiny $ 0.02 $}	
	& $ 0.37 $ & {\tiny $ 0.03 $}	
	& $ \textbf{0.63} $ & {\tiny $ 0.01 $}
	& $ \textbf{0.47} $ & {\tiny $ 0.04 $}
	& $ \textbf{0.69} $ & {\tiny $ 0.01 $} \\ 
%	\hline
	CNTM w network
	& $ \textbf{0.51} $ & {\tiny $ 0.08 $}	
	& $ 0.66 $ & {\tiny $ 0.02 $}
	& $ \textbf{0.39} $ & {\tiny $ 0.03 $}	
	& $ \textbf{0.63} $ & {\tiny $ 0.02 $}
	& $ 0.46 $ & {\tiny $ 0.02 $}
	& $ \textbf{0.69} $ & {\tiny $ 0.01 $} \\ 
	\bottomrule
	\end{tabular}
	\caption{Comparison of clustering performance of CNTM against 
		PMTLM. The best PMTML results are chosen for comparison, 
		from Table 2 in \cite{ZhuYGM:2013}.
	}
	\label{tbl:clustering_LINQS}
	
\end{table}

\subsection{Author-merging for Semi-supervised Learning}

Author modelling allows topic sharing of multiple documents written 
by the same author. However, there are many authors who have 
authored only a few publications, thus their treatment can be 
problematic. In this section, we experiment on merging these authors 
into groups to improve document clustering. We merge authors who 
have authored less than $\eta$ publications, to clarify, $\eta = 2$ 
means authors who have only a single publication are merged, while 
$\eta = 1$ corresponds to no merging. Additionally, we use the 
category labels for a semi-supervised learning. This is achieved by 
assigning the documents to \textit{dummy authors} represented 
by the category labels, \textit{i.e.}\ the authors are merged into 
groups based on the category labels of their publications.
These groups are now considered the ``authors'' for the documents.

\begin{figure}[t!]
	\begin{center}
	\centerline{\includegraphics[width=0.5\columnwidth]{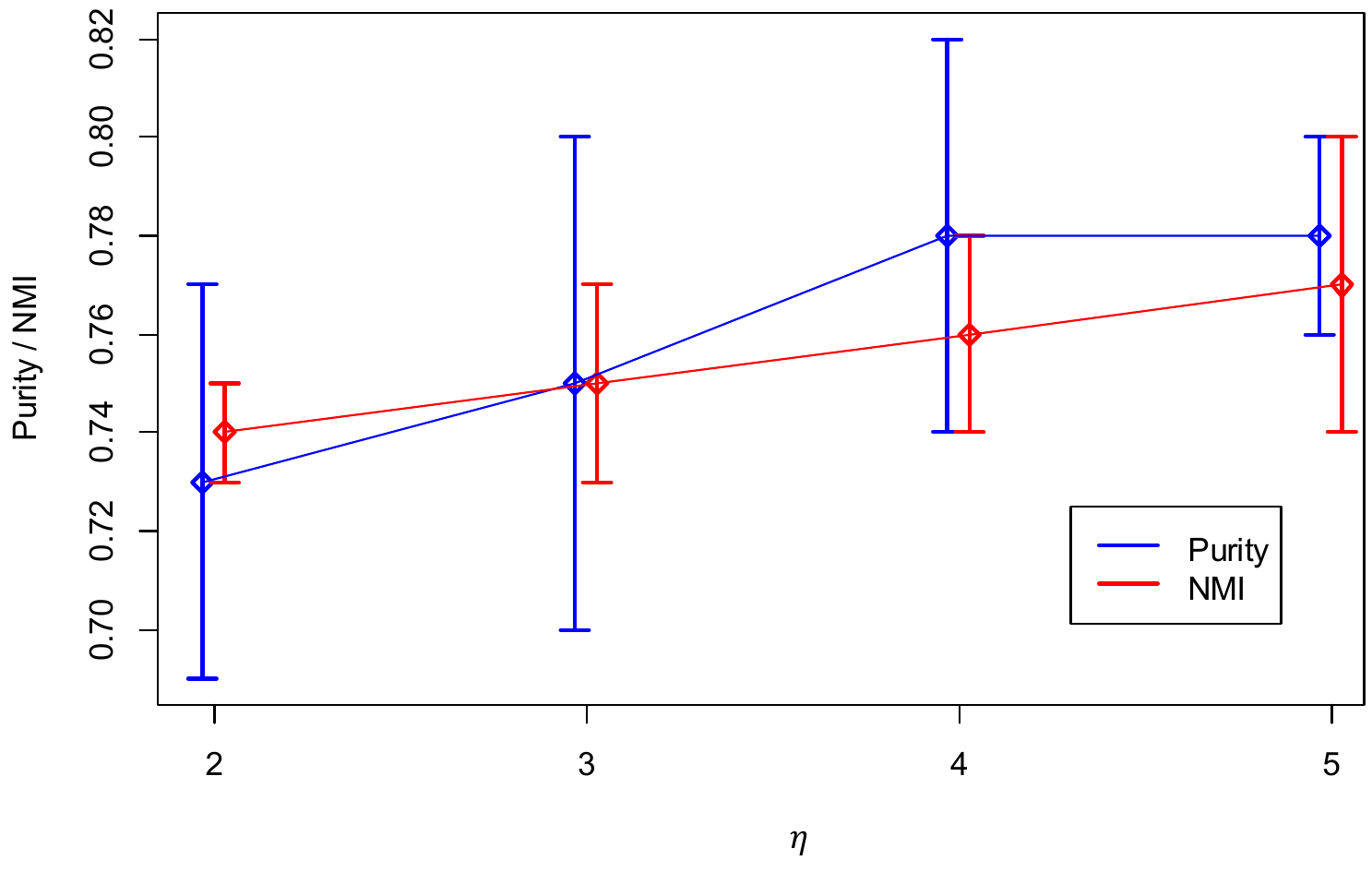}}
	
	\caption{Plot showing the purity and NMI results for 
		$\eta = \{2,3,4,5 \}$ on M10 dataset. The interval 
		represents one standard error for estimation. 
	}
	\label{fig:clustering_m10}
	\end{center}
	
\end{figure}

We present the clustering results for $\eta = \{ 2, 3, 4, 5 \}$ as a 
plot in Figure~\ref{fig:clustering_m10} (results in table format are 
shown in the supplementary material). We find that increasing $\eta$ 
generally improves the clustering performance, although the effect 
is not too significant for successive $\eta$.  Note that if $\eta$ 
is set to be too large, most of the author information will be 
replaced by the category labels, which defeats the purpose of author 
modelling.

\subsection{Qualitative Analysis}

We can obtain a summary of a text corpus from a trained CNTM, this 
is done by analysing the topic-word distribution $\phi$.
In Table~\ref{tbl:topic_ML}, we display some major topics extracted 
from the ML dataset (M10 and AvS in supplementary material).
The topics are represented by the top words, which are ordered based 
on $\phi_{kw}$. The labels of the topics are manually assigned.

\begin{table}[t!]
	\small
	\centering
	\begin{tabular}{cc}
	\toprule
	\multicolumn{1}{c}{Topic} & \multicolumn{1}{c}{Top Words} \\
	\midrule
	Reinforcement Learning & reinforcement, agents, control, state, task \\
%	\hline
	Object Recognition & face, video, object, motion, tracking \\
%	\hline
	Data Mining & mining, data mining, research, patterns, knowledge \\
%	\hline
	SVM & kernel, support vector, training, clustering, space \\
%	\hline
	Speech Recognition & recognition, speech, speech recognition, audio, hidden markov \\
	\bottomrule
	\end{tabular}
	\caption{Topic Summary for ML Dataset}
	\label{tbl:topic_ML}
	
\end{table}

Additionally, we analyse the author-topic distributions $\nu$ to 
find out about authors' interests. We focus on the M10 dataset since 
it covers a wider area of research topics. For each author $a$, we 
determine their dominant topic from their author-topic distribution 
$\nu_a$. 

We display the interests of some authors in 
Table~\ref{tbl:author_interest}. Again, the topic labels are 
manually picked given the dominant topics and the corresponding 
top words from the topics.

\begin{table}[t!]
	\small
	\centering
	
	\begin{tabular}{lcc}
	\toprule
	\multicolumn{1}{c}{Author}	& \multicolumn{1}{c}{Topic} & \multicolumn{1}{c}{Top Words} \\
	\midrule
	D. Aerts & Quantum Theory & quantum, theory, quantum mechanics, classical, quantum field\!\! \\
%	\hline
	Y. Bengio & Neural Network & networks, learning, recurrent neural, neural networks, models \\
%	\hline
	C. Boutilier & Decision Making & decision making, agents, decision, theory, agent \\
%	\hline
	S. Thrun & Robot Learning & robot, robots, control, autonomous, learning \\
%	\hline
	M. Baker & Financial Market & market, risk, firms, returns, financial \\
	\bottomrule
	\end{tabular}
	
	\caption{Example of authors and their topic preference learned by the CNTM.}
	\label{tbl:author_interest}
	\vspace{-1.5mm}
\end{table}

Furthermore, we can graphically visualise the author-topics network 
extracted by CNTM with 
\texttt{Graphviz}.\footnote{\url{http://www.graphviz.org/}}
This is detailed in the supplementary material due to space.

\section{Conclusions}
\label{sec:conclusion}

In this paper, we have proposed the Citation-Network Topic Model 
(CNTM) to jointly model research publications and their citation 
network. CNTM performs text modelling with a hierarchical PYP topic 
model and models the citations with the Poisson distribution.
We also proposed a novel learning algorithm for the CNTM, which 
exploits the conjugacy of the Dirichlet and Multinomial 
distribution, allowing the sampling of the citation networks to be 
of similar form of the collapsed Gibbs sampler of a topic model.
As discussed, our learning algorithm is intuitive and easy to 
implement.

The CNTM offers substantial performance improvement over previous 
work \citep{ZhuYGM:2013}. On three CiteSeer$^\mathrm{X}$ datasets 
and three existing datasets, we demonstrate the improvement of joint 
topic and network modelling in terms of model fitting and clustering 
evaluation. Additionally, we experiment on merging authors who do 
not have many publications into groups of similar authors based on 
the query categories, giving us a semi-supervised learning. We find 
that clustering performance improves with the level of merging.

Future work includes learning the influences of the co-authors, 
utilising them for author merging and further speed up 
non-parametric modelling with techniques in 
\cite{Li:2014:RSC:2623330.2623756}.

\acks{NICTA is funded by the Australian Government through the 
Department of Communications and the Australian Research Council 
through the ICT Centre of Excellence Program. The authors wish to 
thank CiteSeer$^\mathrm{X}$ for providing the data.}

\bibliography{biblio/biblio}

\begin{thebibliography}{27}
\providecommand{\natexlab}[1]{#1}
\providecommand{\url}[1]{\texttt{#1}}
\expandafter\ifx\csname urlstyle\endcsname\relax
  \providecommand{\doi}[1]{doi: #1}\else
  \providecommand{\doi}{doi: \begingroup \urlstyle{rm}\Url}\fi

\bibitem[Buntine and Hutter(2012)]{buntine2012bayesian}
W.~Buntine and M.~Hutter.
\newblock A {B}ayesian view of the {P}oisson-{D}irichlet process.
\newblock Technical Report arXiv:1007.0296v2, 2012.

\bibitem[Buntine and Mishra(2014)]{Buntine:2014:ENT:2623330.2623691}
W.~Buntine and S.~Mishra.
\newblock Experiments with non-parametric topic models.
\newblock In \emph{KDD}, pages 881--890. ACM, 2014.

\bibitem[Chang and Blei(2010)]{chang2010hierarchical}
J.~Chang and D.~Blei.
\newblock Hierarchical relational models for document networks.
\newblock \emph{The Annals of Applied Statistics}, 4\penalty0 (1):\penalty0
  124--150, 2010.

\bibitem[Chen et~al.(2011)Chen, Du, and Buntine]{Chen:2011:STC:2034063.2034095}
C.~Chen, L.~Du, and W.~Buntine.
\newblock Sampling table configurations for the hierarchical
  {P}oisson-{D}irichlet {P}rocess.
\newblock In \emph{ECML}, pages 296--311. Springer-Verlag, 2011.

\bibitem[Han et~al.(2004)Han, Giles, Zha, Li, and
  Tsioutsiouliklis]{Han:2004:TSL:996350.996419}
H.~Han, C.~L. Giles, H.~Zha, C.~Li, and K.~Tsioutsiouliklis.
\newblock Two supervised learning approaches for name disambiguation in author
  citations.
\newblock In \emph{JCDL}, pages 296--305. ACM, 2004.

\bibitem[Han et~al.(2005)Han, Zha, and Giles]{Han:2005:NDA:1065385.1065462}
H.~Han, H.~Zha, and C.~L. Giles.
\newblock Name disambiguation in author citations using a {K}-way spectral
  clustering method.
\newblock In \emph{JCDL}, pages 334--343. ACM, 2005.

\bibitem[Kataria et~al.(2011)Kataria, Mitra, Caragea, and
  Giles]{Kataria:2011:CST:2283696.2283777}
S.~Kataria, P.~Mitra, C.~Caragea, and C.~L. Giles.
\newblock Context sensitive topic models for author influence in document
  networks.
\newblock In \emph{IJCAI}, pages 2274--2280. AAAI Press, 2011.

\bibitem[Li et~al.(2014)Li, Ahmed, Ravi, and
  Smola]{Li:2014:RSC:2623330.2623756}
A.~Li, A.~Ahmed, S.~Ravi, and A.~Smola.
\newblock Reducing the sampling complexity of topic models.
\newblock In \emph{KDD}, pages 891--900. ACM, 2014.

\bibitem[Lim et~al.(2013)Lim, Chen, and Buntine]{Lim2013Twitter}
K.~W. Lim, C.~Chen, and W.~Buntine.
\newblock Twitter-network topic model: A full {B}ayesian treatment for social
  network and text modeling.
\newblock In \emph{NIPS Topic Model workshop}, 2013.

\bibitem[Liu et~al.(2010)Liu, Tang, Han, Jiang, and
  Yang]{Liu:2010:MTI:1871437.1871467}
L.~Liu, J.~Tang, J.~Han, M.~Jiang, and S.~Yang.
\newblock Mining topic-level influence in heterogeneous networks.
\newblock In \emph{CIKM}, pages 199--208. ACM, 2010.

\bibitem[Liu et~al.(2009)Liu, Niculescu-Mizil, and
  Gryc]{Liu:2009:TLJ:1553374.1553460}
Y.~Liu, A.~Niculescu-Mizil, and W.~Gryc.
\newblock Topic-link {LDA}: Joint models of topic and author community.
\newblock In \emph{ICML}, pages 665--672. ACM, 2009.

\bibitem[Lui and Baldwin(2012)]{Lui:2012:LOL:2390470.2390475}
M.~Lui and T.~Baldwin.
\newblock langid.py: An off-the-shelf language identification tool.
\newblock In \emph{ACL}, pages 25--30. ACL, 2012.

\bibitem[Manning et~al.(2008)Manning, Raghavan, and
  Sch\"{u}tze]{Manning:2008:IIR:1394399}
C.~Manning, P.~Raghavan, and H.~Sch\"{u}tze.
\newblock \emph{Introduction to Information Retrieval}.
\newblock Cambridge University Press, 2008.
\newblock ISBN 0521865719, 9780521865715.

\bibitem[Mimno and McCallum(2007)]{Mimno:2007:MDL:1255175.1255196}
D.~Mimno and A.~McCallum.
\newblock Mining a digital library for influential authors.
\newblock In \emph{JCDL}, pages 105--106. ACM, 2007.

\bibitem[Nallapati et~al.(2008)Nallapati, Ahmed, Xing, and
  Cohen]{Nallapati:2008:JLT:1401890.1401957}
R.~Nallapati, A.~Ahmed, E.~Xing, and W.~Cohen.
\newblock Joint latent topic models for text and citations.
\newblock In \emph{KDD}, pages 542--550. ACM, 2008.

\bibitem[Pitman(1996)]{pitman1996some}
J.~Pitman.
\newblock Some developments of the {B}lackwell-{M}acqueen urn scheme.
\newblock \emph{Lecture Notes-Monograph Series}, pages 245--267, 1996.

\bibitem[Rosen-Zvi et~al.(2004)Rosen-Zvi, Griffiths, Steyvers, and
  Smyth]{Rosen-Zvi:2004:AMA:1036843.1036902}
M.~Rosen-Zvi, T.~Griffiths, M.~Steyvers, and P.~Smyth.
\newblock The author-topic model for authors and documents.
\newblock In \emph{UAI}, pages 487--494. AUAI Press, 2004.

\bibitem[Sato and Nakagawa(2010)]{Sato:2010:TMP:1835804.1835890}
I.~Sato and H.~Nakagawa.
\newblock Topic models with power-law using {P}itman-{Y}or process.
\newblock In \emph{KDD}, pages 673--682. ACM, 2010.

\bibitem[Sen et~al.(2008)Sen, Namata, Bilgic, Getoor, Gallagher, and
  Eliassi-Rad]{sen:aimag08}
P.~Sen, G.~Namata, M.~Bilgic, L.~Getoor, B.~Gallagher, and T.~Eliassi-Rad.
\newblock Collective classification in network data.
\newblock \emph{AI Magazine}, 29\penalty0 (3):\penalty0 93--106, 2008.

\bibitem[Tang et~al.(2009)Tang, Sun, Wang, and
  Yang]{Tang:2009:SIA:1557019.1557108}
J.~Tang, J.~Sun, C.~Wang, and Z.~Yang.
\newblock Social influence analysis in large-scale networks.
\newblock In \emph{KDD}, pages 807--816. ACM, 2009.

\bibitem[Teh(2006{\natexlab{a}})]{Teh06abayesian}
Y.~W. Teh.
\newblock A {B}ayesian interpretation of interpolated {K}neser-{N}ey.
\newblock Technical report, School of Computing, National University of
  Singapore, 2006{\natexlab{a}}.

\bibitem[Teh(2006{\natexlab{b}})]{Teh:2006:HBL:1220175.1220299}
Y.~W. Teh.
\newblock A hierarchical {B}ayesian language model based on {P}itman-{Y}or
  processes.
\newblock In \emph{ACL}, pages 985--992. ACL, 2006{\natexlab{b}}.

\bibitem[Teh and Jordan(2010)]{TehJor2010a}
Y.~W. Teh and M.~Jordan.
\newblock Hierarchical {B}ayesian nonparametric models with applications.
\newblock In \emph{Bayesian Nonparametrics: Principles and Practice}. Cambridge
  University Press, 2010.

\bibitem[Tu et~al.(2010)Tu, Johri, Roth, and
  Hockenmaier]{Tu:2010:CAT:1944566.1944711}
Y.~Tu, N.~Johri, D.~Roth, and J.~Hockenmaier.
\newblock Citation author topic model in expert search.
\newblock COLING, pages 1265--1273. ACL, 2010.

\bibitem[Wallach et~al.(2009)Wallach, Mimno, and McCallum]{WallachPrior2009}
H.~Wallach, D.~Mimno, and A.~McCallum.
\newblock Rethinking {LDA}: Why priors matter.
\newblock In \emph{NIPS}, pages 1973--1981. 2009.

\bibitem[Weng et~al.(2010)Weng, Lim, Jiang, and
  He]{Weng:2010:TFT:1718487.1718520}
J.~Weng, E.-P. Lim, J.~Jiang, and Q.~He.
\newblock Twitter{R}ank: Finding topic-sensitive influential {T}witterers.
\newblock In \emph{WSDM}, pages 261--270. ACM, 2010.

\bibitem[Zhu et~al.(2013)Zhu, Yan, Getoor, and Moore]{ZhuYGM:2013}
Y.~Zhu, X.~Yan, L.~Getoor, and C.~Moore.
\newblock Scalable text and link analysis with mixed-topic link models.
\newblock In \emph{KDD}, pages 473--481. ACM, 2013.

\end{thebibliography}

\end{document}